\documentclass[11pt,american]{article}
\usepackage[T1]{fontenc}
\usepackage[utf8]{inputenc}
\usepackage[a4paper]{geometry}
\usepackage{amsmath}
\usepackage{graphicx}
\usepackage{esint}
\usepackage{babel}

\global\long\def\vec#1{\boldsymbol{#1}}
\global\long\def\ten#1{\boldsymbol{#1}}

\global\long\def\parc#1#2{\frac{\partial#1}{\partial#2}}

\global\long\def\ppparc#1#2#3{\frac{\partial^{2}#1}{\partial#2\partial#3}}

\begin{document}
\begin{center}
\textbf{\Large On the compatibility conditions of finite deformations\vspace{12pt}}
\par\end{center}

\begin{center}
\emph{Bal\'azs PERE, PhD, Associate Professor}\textbf{\Large \vspace{12pt}}
\par\end{center}

\begin{center}
{\small Széchenyi István University, 9026 Győr, Egyetem tér 1. (Hungary), e-mail: pere.balazs@sze.hu}\textbf{\Large \vspace{20pt}}
\par\end{center}

\textbf{Abstract}

\emph{This paper examines the intuitive meaning of the Saint--Venant compatibility equation known from the linear theory of deformations. The linearized theory is typically obtained from the relations of finite deformation theory by neglecting higher-order terms. We show how to formulate the compatibility conditions for finite deformations and how the well-known equation for small deformations follows from this formulation.}\textbf{\Large \vspace{11pt}}

\textbf{Keywords}

finite deformations, compatibility condition\textbf{\Large \vspace{11pt}}

\section{Introduction}

In analytical and numerical solutions to linear elasticity problems, the displacement method is used in most cases. In this method, the primary unknown is the displacement field. The remaining unknowns, such as the strain or stress tensor fields, are derived from the displacement vector field through differentiation and algebraic operations. No prior constraint is imposed on the displacement field: given an arbitrary displacement field, both the strain and stress states are uniquely determined. Another classical approach to solving elasticity problems is the force method, in which the stresses are determined first. The difficulty here lies, among other things, in ensuring that the strain field obtained from the stress field satisfies the compatibility condition; otherwise, the displacement field is not uniquely determined \cite{KozakBeda1987,kozak2000}.

In the theory of small deformations, the compatibility condition is given by the Saint--Venant compatibility equation
\begin{equation}
\nabla\times\ten A\times\nabla=\ten 0\label{eq:saint-venant}
\end{equation}
where $\ten A$ denotes the strain tensor describing the strain state at an arbitrary point of the body \cite{KozakBeda1987}. In mechanics, as in other natural sciences that use mathematics, relations and equations carry physical meaning and can often be derived from more fundamental laws. What is the physical meaning of equation \eqref{eq:saint-venant}? Among the relatively small number of publications on this topic, only a few offer such an explanation \cite{kozak1993,haupt2002}. This paper aims to explore the question and provide an intuitive answer.

\section{Formulation of the compatibility condition}

\begin{figure}[h]
\begin{centering}
\includegraphics[scale=0.95]{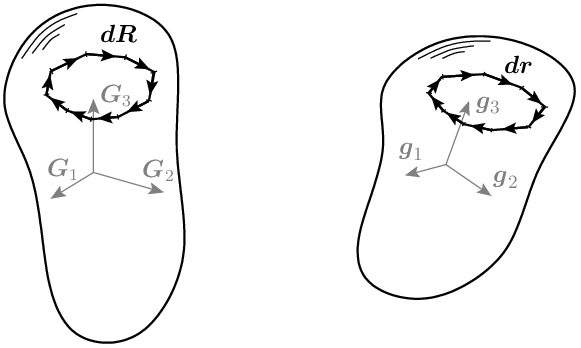}\caption{\label{fig:korint}On both the undeformed and deformed bodies, the sum of elementary vectors along a closed curve is zero.}
\par\end{centering}
\end{figure}

Consider a deformable body in Euclidean space. At a given time $t_{0}$, let $V$ denote the volume of a simply connected subdomain of the body. The configuration of the body at time $t_{0}$ is called the \emph{initial configuration}. The position vector of an arbitrary point in $V$ can be written as
\[
\vec R=X^{I}\vec G_{I}
\]
where $X^{I}$ are the contravariant curvilinear coordinates of the vector and $\vec G_{I}$ are the basis vectors. We adopt Einstein’s summation convention \cite{JanossyTasnadi1982}, i.e., indices repeated twice are summed from one to three. Capital indices indicate that both the coordinates and the basis vectors are defined in the initial configuration. Let $A$ and $B$ be two points in the volume $V$ joined by an arbitrary (smooth) space curve. The vector $\vec R_{AB}$ pointing from $A$ to $B$ can be obtained as the vector sum of the oriented elementary segments of the curve and can be expressed by the line integral
\begin{equation}
\vec R_{AB}=\intop_{A}^{B}d\vec R\label{eq:AB_integral}
\end{equation}
It is straightforward that the vector $\vec R_{AB}$ is independent of the shape and position of the space curve connecting $A$ and $B$. Consider two arbitrary curves $L_{1}$ and $L_{2}$ joining $A$ and $B$. If we compute $\vec R_{AB}$ along curve $L_{1}$ and $\vec R_{BA}$ along curve $L_{2}$ according to relation \eqref{eq:AB_integral}, then the sum of the two vectors must be zero, i.e.,
\begin{equation}
\vec R_{AB}+\vec R_{BA}=\intop_{A}^{B}d\vec R+\intop_{B}^{A}d\vec R=\ointop_{\left(L\right)}d\vec R=\vec 0,\label{eq:korint_dR}
\end{equation}
where $L$ denotes the closed space curve $L_{1}\cup L_{2}$ (see Fig.~\ref{fig:korint}).

\noindent

In the equations above, the vectors are defined in the initial configuration. Now deform the body so that the initial volume $V$ maps to a volume $v$ at a time $t_{1}>t_{0}$. The state corresponding to time $t_{1}$ is called the \emph{current configuration}. To distinguish notationally, we denote quantities defined in the initial configuration by capital letters, while those defined in the current configuration by lowercase letters. Accordingly, the elementary vector $d\vec R$ becomes $d\vec r$, and, following the motion of the material points, the curve $L$ maps to a curve $l$ such that both $L$ and $l$ pass through the same material points. Thus, relation \eqref{eq:korint_dR} written in the initial configuration becomes, in the current configuration,
\begin{equation}
\ointop_{\left(l\right)}d\vec r=\vec 0\label{eq:korint_dr}
\end{equation}
During the deformation, every point initially at $\vec R$ moves to a new position given by the mapping $\vec r=\vec r\left(\vec R\right)$. Expanding the vector-valued function $\vec r\left(\vec R\right)$ in series and retaining the linear terms, equation \eqref{eq:korint_dr} transforms into
\begin{equation}
\ointop_{\left(l\right)}d\vec r=\ointop_{\left(L\right)}\parc{\vec r\left(\vec R\right)}{\vec R}\cdot d\vec R=\vec 0.\label{eq:r_teljes_differencialja}
\end{equation}
The partial derivative on the right-hand side of equation \eqref{eq:r_teljes_differencialja} is the deformation gradient. For clarity in later calculations we use the notation
\[
\ten F=\parc{\vec r\left(\vec R\right)}{\vec R}=\vec r\left(\vec R\right)\circ\nabla_{0}
\]
where $\nabla_{0}$ denotes the nabla operator defined in the initial configuration, and “$\circ$” denotes the dyadic (tensor) product. We shall see that equation \eqref{eq:korint_dr} is, in fact, the compatibility equation for finite deformations, and that by neglecting higher-order terms the compatibility equation \eqref{eq:saint-venant} of the linearized theory is recovered.

\section{Compatibility and the Christoffel–Riemann curvature tensor}

Equation \eqref{eq:korint_dr} (or equivalently \eqref{eq:r_teljes_differencialja}) can be regarded as the integral form of the compatibility equation. Applying Stokes’ theorem \cite{JanossyTasnadi1983},
\begin{equation}
\ointop_{\left(l\right)}d\vec r=\ointop_{\left(L\right)}\vec r\left(\vec R\right)\circ\nabla_{0}\cdot d\vec R=-\intop_{\left(A\right)}\vec r\left(\vec R\right)\circ\nabla_{0}\times\nabla_{0}\cdot d\vec A=\vec 0,\label{eq:Stokes}
\end{equation}
where $A$ is an arbitrary surface with boundary $L$ lying in the volume $V$. Let us transform the integrand obtained after applying Stokes’ theorem in equation \eqref{eq:Stokes}. First, we write the differential operators explicitly:
\[
\vec r\left(\vec R\right)\circ\nabla_{0}\times\nabla_{0}=\parc{}{X^{K}}\left(\parc{}{X^{J}}\left(x^{i}\vec g_{i}\right)\circ\vec G^{J}\right)\times\vec G^{K}=\ten 0,
\]
where $\vec r=x^{i}\vec g_{i}$ is the position vector of material points in the deformed configuration, $x^{i}$ are the contravariant coordinates of $\vec r$, and $\vec g_{i}$ are the basis vectors\footnote{In general $\vec g_{i}\neq\ten F\cdot\vec G_{I}$, since $\vec g_{i}$ and $\vec G_{I}$ are independent bases.}. When performing the indicated differentiations, note that the basis vectors of a curvilinear coordinate system may depend on position; therefore they must also be differentiated with respect to the coordinates $X^{K}$. Carrying out the differentiations yields
\[
\left[\ppparc{x^{i}}{X^{K}}{X^{J}}\left(\delta_{i}^{m}+x^{l}\Gamma_{li}^{m}\right)+\parc{x^{i}}{X^{J}}\parc{x^{p}}{X^{K}}\left(\Gamma_{ip}^{m}+\Gamma_{pi}^{m}+x^{l}\parc{\Gamma_{li}^{m}}{x^{p}}+x^{l}\Gamma_{li}^{n}\Gamma_{np}^{m}\right)\right]\vec g_{m}\circ\vec G^{J}\times\vec G^{K}+
\]
\begin{equation}
+\parc{x^{i}}{X^{N}}\left(\delta_{i}^{m}+x^{l}\Gamma_{li}^{m}\right)\Gamma_{KJ}^{N}\vec g_{m}\circ\vec G^{J}\times\vec G^{K}=\ten 0\label{eq:r_diad_nabla_kereszt_nabla}
\end{equation}
where $\delta_{i}^{m}$ is the Kronecker delta (equal to one if $i=m$ and zero otherwise). Derivatives of the basis vectors are expressed by the Christoffel symbols of the second kind:
\[
\Gamma_{li}^{m}=\parc{\vec g_{l}}{x^{i}}\cdot\vec g^{m}\qquad\mbox{and}\qquad\Gamma_{KJ}^{N}=\parc{\vec G_{K}}{X^{J}}\cdot\vec G^{N}.
\]
Next, examine the elementary surface vector $d\vec A$ in the right-hand integral of equation \eqref{eq:Stokes}. Let $d\tilde{\vec R}=d\tilde{X}^{I}\vec G_{I}$ and $d\hat{\vec R}=d\hat{X}^{J}\vec G_{J}$ be two tangent vectors of the surface starting from a common point and not collinear. Their vector product is the elementary surface vector $d\vec A$:
\begin{equation}
d\vec A=\left(d\tilde{X}^{I}\vec G_{I}\right)\times\left(d\hat{X}^{J}\vec G_{J}\right)=d\tilde{X}^{I}d\hat{X}^{J}\vec G_{I}\times\vec G_{J}.\label{eq:dA}
\end{equation}
Substituting relations \eqref{eq:r_diad_nabla_kereszt_nabla} and \eqref{eq:dA} into equation \eqref{eq:Stokes} and simplifying as needed yields a new relation for the compatibility condition:
\begin{equation}
-\intop_{\left(A\right)}\vec r\left(\vec R\right)\circ\nabla_{0}\times\nabla_{0}\cdot d\vec A=-\intop_{\left(a\right)}x^{l}\underbrace{\left(\parc{\Gamma_{li}^{m}}{x^{p}}-\parc{\Gamma_{lp}^{m}}{x^{i}}+\Gamma_{li}^{n}\Gamma_{np}^{m}-\Gamma_{lp}^{n}\Gamma_{ni}^{m}\right)}_{R_{lpi}^{m}}\vec g_{m}d\tilde{x}^{i}d\hat{x}^{p}=\vec 0.\label{eq:Ch-R-1}
\end{equation}
Equation \eqref{eq:Ch-R-1} is a vector equation and can be written as three scalar equations. The integration domain $a$ is the image in the current configuration of the surface $A$ appearing in Stokes’ theorem \eqref{eq:Stokes}. The $x^{l}$ are the coordinates of an arbitrary point on the surface $a$, and $R_{lpi}^{m}$ are the components of the Christoffel--Riemann curvature tensor \cite{JanossyTasnadi1982,kozak1993,haupt2002}. Rewrite equation \eqref{eq:Ch-R-1} by attaching the corresponding basis vectors to each coordinate. From this, by linearization and using properties of the metric tensor, one arrives at
\begin{equation}
R_{lpi}^{m}\approx\frac{1}{2}\left(\ppparc{g_{iq}}{x^{l}}{x^{p}}-\ppparc{g_{pq}}{x^{l}}{x^{i}}-\ppparc{g_{li}}{x^{q}}{x^{p}}+\ppparc{g_{lp}}{x^{q}}{x^{i}}\right)g^{qm}=0.\label{eq:gorb_tenzor_linearizalva}
\end{equation}
It can be observed that the parenthesized expression in equation \eqref{eq:gorb_tenzor_linearizalva} changes sign when exchanging indices $i$ and $p$, and likewise $l$ and $q$. Furthermore, when $i=p$ and $l=q$, the expression in parentheses is identically zero. Using these observations, equation \eqref{eq:gorb_tenzor_linearizalva} can be written in the form
\begin{equation}
\nabla\times\ten g\times\nabla=\vec 0\label{eq:linearizalt_kompat_felt}
\end{equation}
where
\[
\ten g=g_{ij}\vec g^{i}\circ\vec g^{j}
\]
is the metric tensor. The metric tensor $\ten g$ is a quantity defined in the current configuration.

Return now to the interpretation of the metric tensor mentioned at the beginning of this section: it determines the distance between two nearby points. Let $d\vec r$ be an elementary vector in the current configuration (deformed body) pointing from a point to a nearby point; its length, i.e., the distance in the current configuration, is $dl$. The square of this distance can be computed as
\[
dl^{2}=d\vec r\cdot\tilde{\ten g}\cdot d\vec r=d\vec R\cdot\ten F^{T}\cdot\tilde{\ten g}\cdot\ten F\cdot d\vec R=d\vec R\cdot\bar{\ten G}\cdot d\vec R
\]
where we used $d\vec r=\ten F\cdot d\vec R$. Here $\bar{\ten G}$ is also a metric tensor; for a vector $d\vec R$ defined in the initial configuration it gives the length $dl$ in the current configuration. If an orthonormal Cartesian coordinate system is used in the current configuration, then $\tilde{\ten g}=\ten I$, where $\ten I$ is the identity tensor. This metric tensor can be expressed in terms of the deformation gradient:
\[
\bar{\ten G}=\ten F^{T}\cdot\tilde{\ten g}\cdot\ten F=\ten F^{T}\cdot\ten F=\ten C,
\]
where $\ten C$ is the right Cauchy--Green deformation tensor. Similarly, the length $dL$ in the undeformed body of an elementary vector $d\vec R$ given in the initial configuration is
\[
dL^{2}=d\vec R\cdot\tilde{\ten G}\cdot d\vec R=d\vec r\cdot\ten F^{-T}\cdot\tilde{\ten G}\cdot\ten F^{-1}\cdot d\vec r=d\vec r\cdot\bar{\ten g}\cdot d\vec r,
\]
where $d\vec R=\ten F^{-1}\cdot d\vec r$ and
\[
\bar{\ten g}=\ten F^{-T}\cdot\tilde{\ten G}\cdot\ten F^{-1}.
\]
If an orthonormal Cartesian coordinate system is used in the initial configuration, then $\tilde{\ten G}=\ten I$ and
\[
\bar{\ten g}=\ten F^{-T}\cdot\tilde{\ten G}\cdot\ten F^{-1}=\ten F^{-T}\cdot\ten F^{-1}=\left(\ten F\cdot\ten F^{T}\right)^{-1}=\ten b^{-1},
\]
where $\ten b$ is the left Cauchy--Green deformation tensor.

Among the metric tensors introduced above, the tensor $\ten g$ appearing in equation \eqref{eq:linearizalt_kompat_felt} corresponds to the right Cauchy--Green deformation tensor, since it describes the geometry of the deformed body (see equation \eqref{eq:r_teljes_differencialja}). On this basis, the compatibility equation \eqref{eq:linearizalt_kompat_felt} becomes
\[
\nabla\times\ten g\times\nabla=\nabla\times\ten C\times\nabla=\ten 0
\]
Using the fact that the derivative of a constant is zero, one may replace the right Cauchy--Green deformation tensor $\ten C$ by the Green--Lagrange strain tensor $\ten E=\frac{1}{2}\left(\ten C-\ten I\right)$:
\[
\nabla\times\frac{1}{2}\left(\ten C-\ten I\right)\times\nabla=\nabla\times\ten E\times\nabla=\ten 0
\]
For small deformations, the Green--Lagrange strain tensor is well approximated by the small-strain tensor $\ten A$, i.e., $\ten E\approx\ten A$ \cite{KozakBeda1987}. Thus we recover the Saint--Venant compatibility equation valid in the linear theory of small deformations:
\[
\nabla\times\ten A\times\nabla=\ten 0
\]

\section{Summary}

This paper presented an intuitive interpretation of the compatibility conditions. For finite deformations, the global form of the compatibility condition can be formulated by means of a line integral taken along a closed curve. Using this, one can show that the Christoffel--Riemann curvature tensor corresponding to the deformed geometry of the body is zero at every point of the body. By neglecting higher-order terms, one recovers the Saint--Venant compatibility condition of the linearized theory of deformations.

\end{document}